\documentclass[onecolumn,11pt]{revtex4}
\usepackage{dcolumn}% Align table columns on decimal point
\usepackage{bm}% bold math
\usepackage{amsmath}
\usepackage{amssymb}
\usepackage{revsymb}
\usepackage{graphics}
\usepackage{graphicx}
\usepackage{color}
\usepackage{wrapfig}
\begin{document}

\title{Review of fluctuational electrodynamics and its applications to radiative momentum, energy and entropy transport}% Force line breaks with \\
\author{Yi Zheng}
\email{zheng@egr.uri.edu}
\affiliation{ 
Department of Mechanical, Industrial and Systems Engineering, University of Rhode Island, Kingston, RI 02879, USA
}

\begin{abstract}
Quantum and thermal fluctuations of electromagnetic fields, which give rise to Planck's law of blackbody radiation, are also responsible for van der Waals and Casimir forces, as well as near-field radiative energy transfer between objects. Electromagnetic waves transport energy, momentum, and entropy. For classical thermal radiation, the dependence of the above mentioned quantities on the temperature is well-known mainly due to Planck's work. When near-field effects, namely the collective influence of diffraction, interference, and tunneling of waves, become important, Planck's theory is no longer valid. Of momentum, energy, and entropy transfer, the role of near-field effects on momentum transfer between two half-spaces separated by a vacuum gap (van der Waals pressure in the vacuum gap) was first determined by Lifshitz, using Rytov's theory of fluctuational electrodynamics in 1956. Subsequently, Dzyaloshinskii, Lifshitz, and Pitaevskii, employing sophisticated methods from quantum field theory, generalized Lifshitz' result for van der Waals pressure in a vacuum layer to the case of van der Waals pressure in a dissipative layer between two half-spaces. The influence of near-field effects on radiative transfer was appreciated only in the late 1960s  and, subsequently, in the last two decades because of the enhancement in radiative transfer due to electromagnetic surface waves. The role played by near-field effects on entropy transfer has not been investigated so far, at least when the temperature distribution is non-uniform. 
\end{abstract}

\maketitle

\section{Introduction}

With the development of micro and nano technologies, small scale devices and micro/nanostructured materials find use in different scientific and technological applications.
%Micro and nano technologies have been developing rapidly in the past decades. Small-scale devices and micro/nanostructured materials play a significant role in applications to mechanical, electrical, chemical engineering, biology, and material science. 
The decrease in size not only reduces the usage, cost and demand for emerging materials, but also improves the ability, mobility and sensitivity of instruments, such as sensors and detectors. Meanwhile, it has driven us to further explore advanced science and technology of the micro/nano world. 

One of the many ways in which the small scale world differs from the macroscopic world is the importance of interfacial or surface forces (capillary, surface tension, etc.), which play a significant role in adhesion and cohesion in liquids. In hydrocarbons and non-polar liquids, van der Waals interactions, which are due to the fluctuations of electromagnetic fields, are the main contributions to adhesion and cohesion. The phenomena of adhesion and cohesion play a very important role in engineering, science and technology. It can be used to measure the deformation of carbon nanotube  \cite{hertel1998deformation} and to study the adhesion and wear of micromachined surfaces \cite{delrio2005role,hori2010bacterial}. They are responsible for stiction failure, friction, and some aspect of microfluidics in MEMS/NEMS devices. \cite{delrio2005role,andersson1998van,riddle1990spectral,stifter2000theoretical,dong2010characterization,curtis1970collision,guo2004influence}.  The methods of calculating dispersion force and interaction energy between objects of various geometries can be referred to Ref. \cite{tadmor2001london,bowen1995calculation}. One interesting example is the strong adhesion in gecko setae, one of the most effective adhesives known, which is ascribed to the van der Waals interactions \cite{autumn2002evidence}. Understanding the adhesive phenomenon on a gecko's foot would contribute to applications of biomimicry - designing and manufacturing adhesive microstructures.

%One may wonder ``what will be the end of investigation of emerging small-size technology?" The answer is quite simple and straightforward: there is no such end. As Richard Feynman once said, in the 1960s, ``There's plenty of room at the bottom.” Three or four decades earlier, scientists and researchers were interested in the \textit{micro} stuff. Then break through industrial products came out, for example, the microchip and microprocessor. Nowadays, scientists and researchers are very keen on the materials and structures relevant to the keyword of \textit{nano}, such as nanotube, nanowire, nanoparticle. Then, what will be the dimension and size on which next generation of technology will rely? \textit{Pico-} or \textit{Femto-} scale is not a last stop but an intermediate stop, I firmly believe.

Electromagnetic waves transport energy, momentum, and entropy. For classical thermal radiation, the dependence of the above mentioned quantities on the temperature is well-known due to the works of, mainly, Planck \cite{planck2011theory}. 
%This thesis is focused on research in the area of thermal science and micro/nano-scale heat transfer. It seeks to discover the nature of enhanced thermal transport phenomena due to near-field effects as well as small-scale structure features, and addresses pressing problems involving thermodynamic analysis of MEMS/NEMS and energy research. 
At small scales, the presence of near-field effects, such as interference, diffraction and tunneling of surface waves, dramatically affects thermal transport, which creates abundant phenomena for us to explore. 

The work will focus on small-scale momentum, energy, and entropy transfer (mainly on momentum) via electromagnetic waves due to thermal and quantum fluctuations. A dyadic Green's function formalism is developed to determine near-field radiative energy and momentum transfer between objects of arbitrary shapes and sizes. Momentum transfer due to electromagnetic fluctuations is responsible for van der Waals and Casimir forces, which are important in many different fields such as adhesion and stiction of materials, bioengineering, and phase change heat transfer. I show how my work provides a new interpretation, and a better understanding, of this phenomenon in Ref. \cite{zheng2011lifshitz,narayanaswamy2013van}. For energy transfer, I wanted to expose the key differences between momentum and energy transfer. For the last few years, we (others in Narayanaswamy group, and other groups) have been trying to understand the similarities between energy and momentum transfer. However, in doing so, I wanted to solve a slightely more challenging, and hopefully more useful, problem of contributions to the van der Waals pressure and radiative transfer at a point on a half-space from different parts of the second half-space \cite{zheng2014patch}. Thus, we showed the differences between energy and momentum transfer. This work was made possible only because of the dyadic Green's function based surface intergral formalism that I developed in Ref. \cite{narayanaswamy2013green}. An attempt to understand van der Waals forces in a thermal non-equilibrium condition, i.e., when the objects are at unequal temperatures, led us to the analysis of entropy transfer between two half-spaces when near-field effects are present. We used the entropy transfer formulism we developed in Ref. \cite{narayanaswamy2013theory} to also find the maximum work that can be extracted in near-field thermal radiation. However, we have so far been unsuccessful in applying one formalism to the problem we originally intended to solve, i.e., that of van der Waals forces in thermal non-equilibrium condition.

 %In addition, entropy associated with near-field radiative transfer has been studied for the first time. It can be used to determine the maximum work that can be extracted and a thermodynamic limit of energy conversion efficiency that can be obtained in near-field thermal radiation. Small-scale thermal transport has shown great potential and applications for use in manipulating macroscale energy systems and energy harvesting. %Future research work will focus on the thermal and optical properties of 2D or 3D nanostructured materials, and will lead to new types of thermophotovoltaic solar cells and selective thermal emitters using meta-materials and nanoparticles \cite{sai2001spectral,sergeant2010high,yeng2012enabling}. 

%Now, let us think about \textit{how to study the thermal transport at small scale?} One effective and necessary approach is to investigate the behavior of fluctuating charges,

To do all this, we need to have a basic understanding of fluctuations of charges due to temperature as well as quantum effects. That is provided by Rytov's theory of fluctuational electrodynamics \cite{rytov1967theory,stratton2007electromagnetic}, which leads to thermal radiative heat transfer and was also the basic for Lifshitz' celebrated work on van der Waals force. 
%Understanding how charges fluctuate and interact within and between condensed phases (liquids and solids) of a material will provide us an elaborate view about the fluctuating electric and magnetic field \cite{rytov1967theory,stratton2007electromagnetic}, which lead to thermal radiative heat transfer and fluctuation induced forces. 
In the 1950s, Rytov established the fluctuation-dissipation theorem, which can be  thought of as a combination of statistical physics, quantum physics, and macroscopic electrodynamics \cite{callen1951fluctuation,rytov1967theory}. This well-known theory is used to relate the power spectral density of fluctuating charge density to the local temperature, and frequency dependent relative dielectric permittivity and relative magnetic permeability of an object \cite{callen1951fluctuation}. %The fluctuation of charges and field occur because of both thermal effects and quantum-mechanical uncertainties of the positions and momenta of particles, and they affects electromagnetic interactions at the small scale. 
The essence of the theory of fluctuational electrodynamics is the frequency distribution of the fluctuations and its connection to the dissipation of electromagnetic waves imposed on them \cite{parsegian2006van}. 
%, which leads to the fluctuation induced van der Waals and Casimir forces via momentum transfer \cite{lifshitz1956theory,zheng2011lifshitz}, and of near-field thermal radiation via energy as well as entropy transfer \cite{stratton2007electromagnetic,hu2008near}.

\section{Momentum transfer: van der Waals/Casimir pressure}

The theory of van der Waals force is now applicable to kinds of theoretical and experimental investigations, for various geometries other than parallel plates as previously descibed. During the past half century, many theoretical works were published on the topic of van der Waals forces \cite{de1936influence,nesterenko2012lifshitz,buckingham1988theoretical,karplus2004van,israelachvili2011intermolecular,rodriguez2011casimir,pitaevskii2010casimir,gingell1973prediction,parsegian1973van,munday2009measured,ninham1970van,mclachlan1964van,intravaia2012casimir,schwinger1978casimir,milton1997casimir,li2006london,lenac2008casimir,svetovoy2007evanescent,narayanaswamy2013van,guo2004influence,boyer1968quantum,rahi2009scattering,kruger2012trace}. For example, van der Waals forces between an atom and a flat surface \cite{mclachlan1964van}, van der Waals force in multilayered structures \cite{parsegian1973van}, the influence of van der Waals forces and primary bonds on binding energy and strength of natural and artificial resins \cite{de1936influence}, interfacial Lifshitz/van der Waals and polar interactions between macroscopic objects \cite{van1988interfacial,parsegian1972van,ninham1970macroscopic} and between molecules \cite{buckingham1988theoretical,karplus2004van,israelachvili2011intermolecular,zheng2014radiative}, effects of van der Waals force in chemistry and biology \cite{bartell2004effects}, thermal non-equilibrium Casimir/Lifshitz force \cite{antezza2005new,antezza2008casimir}, London-van der Waals interactions between rough bodies \cite {li2006london}, and Casimir effects Polder, van der Waals/Casimir effect in micro- and nano-structured geometries \cite{rodriguez2011casimir}.

%\subsection{Theory}

The fluctuation of charges and fields gives rise to the van der Waals force  \cite{margenau1939van}. That is the interaction between atom or molecules, making particles of materials compact and congregate to create condensed phases, like liquids and solids. Otherwise, they are sparsely distributed gases \cite{slater1931van,bradley1932lxxix}. When the separation is large compared to the size of particles, there exist dipole-dipole interactions inbetween. The dipole can simply be considered to be a pair of positive and negative charges in a neutral particle. The free energy between these neutral particles, which is the work required to bring them from infinite separation to a finite distance $r$, varies as the inverse sixth power of distance, $C/r^6$, where coefficient $C$ can be attributed to Keesom interactions, Debye interactions, or London dispersion interactions \cite{keesom1912deduction,keesom1915second,debye1920van,london1937general,parsegian1973van}.

The distance $r$ should be much greater than dipole or particle size itself. However, if the separation is comparable to the size of atom or molecule, the van der Waals force dominates the interactions, which follows a power law as $1/r^3$ or $1/r^4$ if the finite speed of light is taken into account \cite{london1937general,hamaker1937london,casimir1946influence,van1968macroscopic,gregory1981approximate}. In 1937, Hamaker \cite{hamaker1937london} investigated the properties of van der Waals interactions between macroscopic bodies, which differ from the interactions between individual particles or molecules that had been studied previously. The method that Hamaker used is called pairwise-summation approximation. It is used to sum the interactive energy of all the dipoles over the entire volumes of the two planar bodies (half-spaces) separated by a vacuum gap $d$, which is smaller than the depth and lateral dimensions of the half-spaces. It was shown that the free energy $C/r^6$  between charges turns out to be an energy per unit area that obeys a power law $\propto 1/d^2$. Similarly using the pair-wise summation, the interaction between two spheres, instead of two planar bodies, of radii $R$ with a separation $d$ (here, $d \ll R$), obeys a power law as the inverse of the separation $R/d$. As expected, it yields an inverse sixth power of distance between two spheres which are separated widely, $1/r^6$ (where $r=d+2R$) \cite{london1937general,london1942centers}. 

Hamaker \cite{hamaker1937london} defined a coefficient, that came to be known later as the \textit{Hamaker coefficient} ($A_H$), to characterize the van der Waals interactions between macroscopic objects. It was used to express the free energy between two planar bodies with a separation of $d$ as $U_{free}=-A_H/12\pi d^2$. Differentiating this free energy with respect to distance gives us the van der Waals pressure as $F_{vdW}=A_H/6\pi d^3$. In Ref. \cite{casimir1946influence}, Casimir focused on the free energy of electromagnetic modes and derived an electromagnetic (later shown as van der Waals) pressure between two ideally conducting metal plates by defining a \textit{zero-point} energy of interaction in terms of the number of electromagnetic modes within a cavity at finite spacing.
%charge motions for two plates infinitely separated. 
The main contribution of Casimir's work was not simply a calculation of van der Waals interaction between two conducting walls, but to broaden the our view of van der Waals interaction from microscopic level to macroscopic level, so that people, such as Lifshitz \cite{lifshitz1956theory}, Dzyaloshinskii\cite {dzyaloshinskii1961general}, Oss \cite{van1988interfacial}, Antezza \cite{antezza2008casimir}, Zheng \cite{zheng2011lifshitz} and many others, would be able to extend the work on van der Waals/Casimir force to objects of arbitrary shapes and sizes.

%Two years later, in 1948, Casimir and Polder \cite{casimir1948influence} showed the effect of retardation on the fluctuating electromagnetic field induced force between particles. The finite time for waves or fields traveling from one charge to the other, needs to be taken into consideration. We can think about it as the respond time for a second charge to respond to the fluctuating field induced by the first charge, which will always weaken the strength of the interaction between the two particles or larger bodies. The direct change is in the distance dependence. While considering the retardation effect, the free energy between particles with a distance $r$ changes from $1/r^6$ to $1/r^7$, and the free energy between two planar plates with a distance $d$ changes from $1/d^2$ to $1/d^3$.

\begin{figure}[h]
\centering
\includegraphics[width =7.0cm]{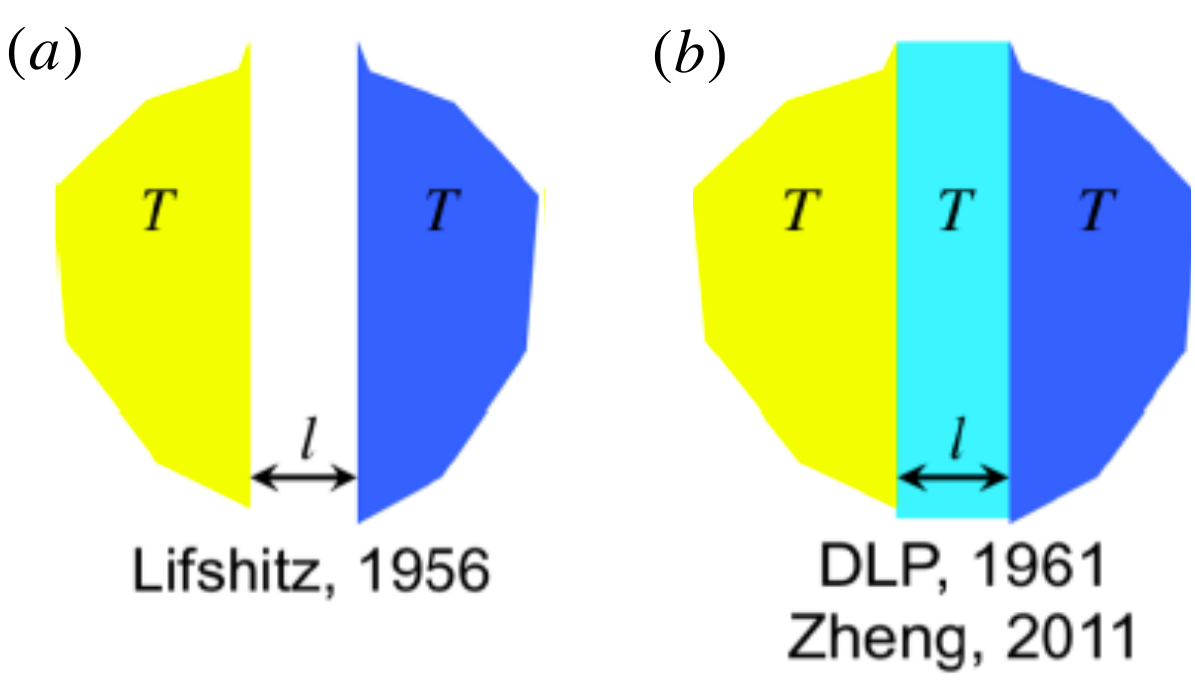}
\caption{\label{fig:eqforce} van der Waals force at thermal equilibrium between two half-spaces at temperature $T$ separated by (a) vacuum gap of $l$, that was solved by Lifshitz in 1956, and (b) a dissipative medium, that was solved by Dzyaloshinskii et al. in 1961, and Zheng and Narayanaswamy in 2011.
}
\end{figure}

In 1950s, Rytov developed the theory of fluctuational electrodynamics for studying the phenomena of fluctuating waves and fields within a material or across an interface between two materials \cite{rytov1967theory}. In 1956, Lifsthiz, based on Rytov's theory of fluctuational electrodynamics, outlined a method in his seminal work \cite{lifshitz1956theory} to calculate the van der Waals/Casimir force between two plates separated in vacuum, as shown in Fig. \ref{fig:eqforce}(a), by defining an electromagnetic stress tensor, which is valid only in vacuum. A few years later, Dzyaloshinskii, Lifshitz and Pitaevskii \cite{dzyaloshinskii1961general} relied on quantum field theory and accomplished a general theory by replacing a vacuum medium by any dissipative medium, as shown in Fig. \ref{fig:eqforce}(b), for which the relative permittivity ($\varepsilon_v=1$) and relative permeability ($\mu_v=1$) in vacuum were simply replaced by $\varepsilon_m(\omega)$ and $\mu_m(\omega)$ of other material. This replacement seems quite simple and easily understood, but its derivation was not as simple. In next subsection, the validity and limitation of Lifshitz' method will be elaborated.

\subsection{\label{sec:whyinvalid}Why is the electromagnetic stress tensor invalid in dissipative media?}

My main motivation for taking on the problem of van der Waals force in dissipative media was simple: why did Dzyaloshinskii take so long after the publication of Lifshitz' work, and so much tougher an approach, for the generalization to dissipative media? The key to that can be found in a recent article by Pitaevskii \cite{pitaevskii2011problem}, outlining some of the history of the work in Landau's group on topic of van der Waals force from the 1960's. The reason was that they did not (we do not even now) have an expression for electromagnetic stress tensor in any media that have dissipative properties as shown in Fig. \ref{fig:eqforce}(b), i.e., when the dieletric function and/or magnetic permeability are frequency dependent and complex. Since we deal with frequency dependent properties, we will be using frequency domain Maxwell's equations and fields as our basis.

To see why that is the case, we can go to a slightly simpler problem - that of energy density in a dissipative medium. Energy density and stress tensor are closely related - both have the same dimensions. The energy flow, also called the Poynting vector, is given by $\bm{S}=\bm{E}\times\bm{H}$, which remains valid even if dispersion is present. It is evident from the condition of the continuity of the tangential components of the electric field $\bm{E}$ and magnetic field $\bm{H}$ at the boundary of the object, that is to say, the normal component of the Poynting vector $\bm{S}$ being continuous at the boundary of the body as well as in the vacuum. 

We know the rate of change of the energy per unit volume is the divergence of the Poynting vector. Using the Ampere's law and Faraday's law in the Maxwell's equations, given by
\begin{subequations}
%\begin{equation}
%\label{eqn:coulomb}
%\text{Coulomb's law: } \nabla\cdot \bm{D}=\rho
%\end{equation}
%\begin{equation}
%\label{eqn:gauss}
%\text{Gauss's law:  } \nabla\cdot \bm{B}=0
%\end{equation}
\begin{equation}
\label{eqn:ampere}
\nabla\times \bm{H}=\bm{J} + \frac{\partial \bm{D}}{\partial t}
\end{equation}
\begin{equation}
\label{eqn:faraday}
\nabla\times \bm{E}=-\frac{\partial \bm{B}}{\partial t}
\end{equation}
\end{subequations}
where $\bm{J}$ is electric current density. $\bm{E}$ and $\bm{B}$ are electric and magnetic fields, $\bm{D}$ and $\bm{H}$ are corresponding derived fields, related to $\bm{E}$ and $\bm{B}$ through the polarization $\bm{P}$ and the magnetization $\bm{M}$ by $\bm{D}=\varepsilon_0 \bm{E}+\bm{P}$ and $\bm{B}=\mu_0 (\bm{H} + \bm{M})$, and $\varepsilon_0$ and $\mu_0$ are the electric permittivity and the magnetic permeability in free space respectively, we can write the rate of change of energy as
\begin{equation}
\begin{split}
\label{eqn:divS}
&\nabla\cdot\bm{S}=\bm{H}\cdot\left(\nabla\times \bm{E}\right)-\bm{E}\cdot\left(\nabla\times \bm{H}\right) -\bm{E}\cdot\bm{J}\\
&\Rightarrow \nabla\cdot\bm{S}+\left(\bm{E}\cdot \frac{\partial \bm{D}}{\partial t} + \bm{H}\cdot \frac{\partial \bm{B}}{\partial t} \right)=-\bm{E}\cdot\bm{J}
\end{split}
\end{equation}
Eq. \ref{eqn:divS} denotes the conversion of energy for a system of electromagnetic fields. The first term on the left $\nabla\cdot\bm{S}$ is the rate of flow of energy across the boundary of object, the sum of two terms $\bm{E}\cdot \bm{D}$ and $\bm{H}\cdot \bm{B}$ on the left represents the total electromagnetic energy density for both static and time-varying fields (Please see Eq. 4.89 and Eq. 5.148 in Ref. \cite{jackson1962classical}),
%the second term $\displaystyle \left(\bm{E}\cdot \frac{\partial \bm{D}}{\partial t} + \bm{H}\cdot \frac{\partial \bm{B}}{\partial t} \right)$ is the rate of change in the internal energy,
and the term on the right $-\bm{E}\cdot\bm{J}$ is the rate of change in the energy of the source due to field. In a non-dissipative medium (free space), when the permittivity $\varepsilon_0$ and the permeability $\mu_0$ are real, the second term on the left can be regarded as the rate of change of the electromagnetic energy, $\displaystyle \frac{\partial u}{\partial t}$, where energy density $u$ given by 
\begin{equation}
\label{eqn:u}
u=\frac{1}{2}\left(\bm{E}\cdot\bm{D}+\bm{H}\cdot\bm{B}\right)=\frac{1}{2}\left(\varepsilon_0|{E}|^2+\mu_0|{H}|^2\right)
\end{equation}
The energy density $u$ in Eq. \ref{eqn:u} has an exact thermodynamic significance that it is the difference of the internal energy per unit volume with and without the electromagnetic field. 

In the presence of the dissipation, the electromagnetic energy cannot be defined as a thermodynamic quantity \cite{zheng2014thermodynamic}, because the energy is dissipated or absorbed in the medium. %Previously, the non-dissipation or lossless relations are $\bm{D}=\varepsilon\bm{E}$ and $\bm{B}=\mu\bm{H}$ with the real and frquency independent $\varepsilon$ and $\mu$. But the 
The actual materials exhibit dispersion or absorption, with complex and frequency dependent properties $\varepsilon(\omega)=\varepsilon'(\omega)+i\varepsilon''(\omega)$ and $\mu(\omega)=\mu'(\omega)+i\mu''(\omega)$. %As a result, the second term on left hand side of Eq. \ref{eqn:divS} cannot be expressed simply as the time derivative of energy density ($\partial u/\partial t$). 
For this, we follow Jackson's work (section 6.8 in Ref. \cite{jackson1962classical}) on Poynting's theorem for electromagnetic fields in linear dissipative media. The assumption of linearity (for simplicity, isotropy is also assumed) implies that $\bm{D}(\bm{r},\omega)=\varepsilon(\omega)\bm{E}(\bm{r},\omega)$ and $\bm{B}(\bm{r},\omega)=\mu(\omega)\bm{H}(\bm{r},\omega)$. I also assume that the $\bm{E}$ and $\bm{H}$ fields are dominated by a relatively narrow range of frequencies, i.e. $\bm{E}=\tilde{\bm{E}}(t)\cos(\omega_0t+\alpha)$, $\bm{H}=\tilde{\bm{H}}(t)\cos(\omega_0t+\beta)$, where $\tilde{\bm{E}}(t)$ and $\tilde{\bm{H}}(t)$ are slowly varying relative to $1/\omega_0$ and the inverse of the frequency range over which $\varepsilon(\omega)$ changes appreciably. The key result of his work for quasi-monochromatic electromagnetic fields is, on averaging with respect to time, given by
\begin{equation}
\begin{split}
\label{eqn:newdivS}
%\left<\bm{E}\cdot \frac{\partial \bm{D}}{\partial t} + \bm{H}\cdot \frac{\partial \bm{B}}{\partial t} \right>= &\omega_0 \varepsilon''(\omega_0) \left<\bm{E}(\bm{r},t)\cdot\bm{E}(\bm{r},t)\right> + \\ & \omega_0 \mu''(\omega_0) \left<\bm{H}(\bm{r},t)\cdot\bm{H}(\bm{r},t)\right>+\frac{\partial u_{\text{eff}}}{\partial t}
\left<\bm{E}\cdot \frac{\partial \bm{D}}{\partial t} + \bm{H}\cdot \frac{\partial \bm{B}}{\partial t} \right>= &\omega_0 \varepsilon''(\omega_0) |{E}|^2 +  \omega_0 \mu''(\omega_0) |{H}|^2+\frac{\partial u_{\text{eff}}}{\partial t}
\end{split}
\end{equation}
where, the $\left< \right>$  means an averaging over the time period of the frequency of the fields. The first two terms on the right %, which are related to the imaginary parts of properties $\varepsilon$ and $\mu$, 
are the conversion of electric and magnetic energy into thermal or mechanical energy, and the second term is the time derivative of the effective energy density, given by 
\cite{brillouin1921propagation,landau1984electrodynamics}
\begin{equation}
\begin{split}
\label{eqn:ueff}
%u_{\text{eff}}(\omega)=&\frac{1}{2}Re\left(\frac{d(\omega \varepsilon(\omega))}{d\omega}\right)\left<\bm{E}(\bm{r},t)\cdot\bm{E}(\bm{r},t)\right>+\\ &\frac{1}{2} Re\left(\frac{d(\omega \mu(\omega))}{d\omega}\right)\left<\bm{H}(\bm{r},t)\cdot\bm{H}(\bm{r},t)\right>
u_{\text{eff}}(\omega)=&\frac{1}{2}Re\left(\frac{d(\omega \varepsilon(\omega))}{d\omega}\right)|{E}|^2+\frac{1}{2} Re\left(\frac{d(\omega \mu(\omega))}{d\omega}\right)|{H}|^2
\end{split}
\end{equation}
Note that, if $\varepsilon$ and $\mu$ are real and frequency independent, Eq. \ref{eqn:newdivS}  gives the simple expression $\partial u/\partial t$, with $\varepsilon''=\mu''=0$ and Eq. \ref{eqn:ueff} recovers the energy density in free space in Eq. \ref{eqn:u}. The change of energy in Eq. \ref{eqn:newdivS} shows that the dissipation or absorption of energy is determined by the imaginary parts of $\varepsilon$ and $\mu$, the first two terms on the right of Eq. \ref{eqn:newdivS}, which are called the electric and magnetic losses respectively. For most substances at positive frequencies, $\varepsilon''>0$ and $\mu''>0$. If there exist very small losses in certain frequency ranges, in which $\varepsilon''$ and $\mu''$ are very small compared with $\varepsilon'$ and $\mu'$, such ranges are called transparency ranges. It is possible to neglect the absortion and use Eq. \ref{eqn:u} to represent the electromagnetic energy density (Please refer to section 80 of Ref. \cite{landau1984electrodynamics}, sections 1.2, 1.4 of Ref. \cite{kong1986electromagnetic} and sections 6.1, 6.6-6.8, 7.6 of Ref. \cite{jackson1962classical}.)

%Let us look at the energy density of an arbitrary electromagnetic field in vacuum. It is simply given by $\displaystyle u=\frac{1}{2}\varepsilon_0|\bm{E}|^2+\frac{1}{2}\mu_0|\bm{H}|^2$. Now let us look at what the energy density would be if the vacuum were replaced with a material with $\varepsilon(\omega)$ and/or $\mu(\omega)$. For this, we follow Jackson's work (section 6.8 in Ref. \cite{jackson1962classical}) on Poynting's Theorem for electromagnetic fields in linear dissipative media. The key result of his work for quasi-monochromatic electromagnetic fields is that the energy density, which is given by
%\cite{brillouin1921propagation,landau1984electrodynamics}
%\begin{equation}
%\label{eqn:ueff}
%u_{\text{eff}}(\omega)=\frac{1}{2}\left[Re\left(\frac{d(\omega \varepsilon(\omega))}{d\omega}\right)\left<\bm{E}\cdot\bm{E}\right>+Re\left(\frac{d(\omega \mu(\omega))}{d\omega}\right)\left<\bm{H}\cdot\bm{H}\right>\right]
%\end{equation}
%where the $\left< \right>$  means an averaging over the time period of the frequency of the fields.

%The velocity of propagation of electromagnetic energy through a non-dissipative media is equal to the group velocity $v_g=\partial \omega/\partial k$. In an absorbting media, the difficuly arises in relating the group velocity to the velocity of energy propagation, because wave vector is complex. 

The question arises as to what happens when $\displaystyle Re\left(\frac{d(\omega \varepsilon)}{d\omega}\right)$ or $\displaystyle Re\left(\frac{d(\omega \mu)}{d\omega}\right)$ is negative? To solve this problem, Loudon \cite{loudon1970propagation,loudon1997propagation}, and later Ruppin \cite{ruppin2002electromagnetic}, used a Lorentz oscillator (mass-spring oscillator) model to arrive at a better expression for energy density, which circumvents the problem of negative energy density. In classical theory, a dispersive and dissipative medium can be described by a collection of damped, non-interacting, harmonic oscillators of displacement ${r}$, mass $M$, natural frequency $\omega_0$, and effective charge $e$. Assuming a damping proportional to $\dot{{r}}$, the equation of motion of an oscillator in the presence of an oscillating electric field is given by \cite{loudon1970propagation,jackson1962classical}
\begin{equation}
\label{eqn:classicalmodel}
M\left({r}+\Gamma{r}+\omega_0^2{r}\right)=e{E} 
\end{equation}
For example, in an absorbing medium, the dielectric function is in the form of 
\begin{equation}
\label{eqn:diefunc}
%\varepsilon(\omega)=(n+i\kappa)^2=\varepsilon_\infty\left(1+\frac{\omega_{TO}^2-\omega_{LO}^2}{\omega^2-\omega_{TO}^2+i\omega\Gamma}\right)
\varepsilon(\omega)=\left[n(\omega)+i\kappa(\omega)\right]^2=\varepsilon_\infty-\frac{4\pi e^2}{MV}\frac{1}{\omega^2-\omega_0^2+i\omega\Gamma}
\end{equation}
where $n$ and $\kappa$ are the real and imaginary parts of refractive index, %$\varepsilon_\infty$ is the background dielectric constant, %$\omega_{TO}$ and $\omega_{LO}$ are the transverse and longitudinal frequencies, 
and $\Gamma$ is the damping constant. It is convenient to define a frequency $\beta$ by $\beta^2=4\pi e^2/MV\varepsilon_\infty$. That is the frequency of plasma oscillations of a collection of charge carriers of mass $M$, charge $e$ and concentration $1/V$ in a medium of background dielectric constant $\varepsilon_{\infty}$. Loudon and Ruppin related the energy density associated with an electromagnetic wave passing through an absorbing dielectric material to the optical properties and the parameters of the model used to describe the absorbing medium, which is given by (Eq. 15 in Ref. \cite{loudon1970propagation})
\begin{equation}
\label{eqn:uloudonEq15}
u_{\text{eff}}(\omega)=\frac{2\pi M}{V}\left(\dot{r}^2+\omega_0^2 r^2\right)+\frac{\varepsilon_\infty |E|^2+|H|^2}{2}
\end{equation}
Using ${H}=(n+i\kappa){E}$ (Please see Eq. 17 in Ref. \cite{loudon1970propagation} and Eq. 7.1 in Ref. \cite{bagrov1990exact}, only valid for an electromagnetic plane wave), Eq. \ref{eqn:classicalmodel} and Eq. \ref{eqn:diefunc} to eliminate $r$ and $H$, Eq. \ref{eqn:uloudonEq15} can be re-written as
\begin{equation}
\label{eqn:uloudon}
%u_{\text{eff}}(\omega)=\frac{\left<\bm{E}(\bm{r},t)\cdot\bm{E}(\bm{r},t)\right>}{2}\left[n(\omega)^2+\frac{2\omega n(\omega) \kappa(\omega)}{\Gamma}\right]
u_{\text{eff}}(\omega)=\frac{|{E}|^2}{2}\left[n(\omega)^2+\frac{2\omega n(\omega) \kappa(\omega)}{\Gamma}\right]
\end{equation}
In the limit of zero damping, where $\Gamma$, $\kappa \rightarrow 0$, the dielectric function can be simplified as
\begin{equation}
\label{eqn:nkzero}
\varepsilon(\omega)=n(\omega)^2=\varepsilon_{\infty}\left(1-\frac{\beta^2}{\omega^2-\omega_0^2}\right)
\end{equation}
Substituting
%use ${H}=(n+i\kappa){E}$ (Please see Eq. 17 in Ref. \cite{loudon1970propagation} and Eq. 7.1 in Ref. \cite{bagrov1990exact}, only valid for an electromagnetic plane wave), 
Eq. \ref{eqn:diefunc}, and Eq. \ref{eqn:nkzero} into Eq. \ref{eqn:ueff} (or Eq. \ref{eqn:uloudon}), we obtain the expression for energy density in a dissipative medium with zero damping (See Eq. 8, Eq. 9, and Eq. 19 in Ref. \cite{loudon1970propagation}, also derived for a special case of a general result of zero damping by others \cite{brillouin1960wave,landau1960classical,pelzer1951energy,ruppin2002electromagnetic})
\begin{equation}
\begin{split}
\label{eqn:ueffzero}
u_{\text{eff}}(\omega)
%=&\frac{1}{2}\left(\varepsilon(\omega)+\omega\frac{d \varepsilon(\omega)}{d\omega}\right)\left<\bm{E}(\bm{r},t)\cdot\bm{E}(\bm{r},t)\right>+\frac{1}{2}\varepsilon(\omega)\left<\bm{E}(\bm{r},t)\cdot\bm{E}(\bm{r},t)\right>\\
%=&\frac{|{E}|^2}{2}\left(\varepsilon(\omega)+\omega\frac{d \varepsilon(\omega)}{d\omega}+\varepsilon(\omega)\right)\\
=&\frac{|{E}|^2}{2}\left[\varepsilon_\infty\left(1+\frac{ \beta^2\left(\omega^2+\omega_0^2\right)}{\left(\omega^2-\omega_0^2\right)^2}\right)+\varepsilon(\omega) \right]\\
=&\frac{|{E}|^2}{2}\left(\frac{\omega}{2}\frac{d\varepsilon(\omega)}{d\omega}+\varepsilon(\omega) \right)\\
=&\frac{|{E}|^2}{2}\left(n(\omega) \omega \frac{dn(\omega)}{d\omega}+n^2 \right)\\
=&\frac{|{E}|^2}{2} n(\omega) \frac{d\left(n(\omega) \omega\right)}{d\omega}
\end{split}
\end{equation}
The effective energy density can also be expressed in terms of the dielectric function of dissipative material instead of the refractive index by Zhang \cite{zhang2007nano,zhang2013measurements}.

%$\displaystyle u_{\text{eff}}(\omega)=\frac{\left<\bm{E}\cdot\bm{E}\right>}{2}n\frac{d(n\omega)}{d\omega}$ (See Eqs. 8, 9, and 19 in Ref. \cite{loudon1970propagation}). 

However, the difficulty they introduce is that instead of a simple dielectric function appearing in their formula, we now have the microscopic parameters that affect $\varepsilon$ appear directly in the formula. Alternatively, one could use a more complicated Navier-Stokes like description for the motion of charges within a fluid and obtain a formula for energy density specific to that model. Does this mean that there is no satisfactory formula for energy density with just $\varepsilon$ and $\mu$ involved?

The electromagnetic stress tensor is no different. The same challenges appear in a different form when we try to find an expression for stress tensor in dissipative media. Of course, we are imposing a tough constraint ourselves - namely, a formula for arbitrary electromagnetic fields. However, we are interested in fields that arise out of thermal and quantum mechanical fluctuation, i.e., fields that obey Rytov's theory of fluctuational electrodynamics. In conclusion, there is no a general form of stress tensor in an absorbing or dissipative medium, in which the tensor cannot be expressed simply in terms of permittivity and permeability alone \cite{landau1984electrodynamics}. If it does exist, it could be considerably complicated in comparison to the well defined Maxwell stress tensor in free space (please refer to section 6.8 of Ref. \cite{jackson1962classical} and sections 10, 15, 35, 75, 80, 81 of Ref. \cite{landau1984electrodynamics}).

During the past few decades, there were several attempts trying to perfect the Maxwell stress tensor for a dispersive medium with losses, that might work for a specific range of frequencies or under a debatable assumption. For instant, Ninham et al. postulated the free energy of an electromagnetic mode at frequency $\displaystyle \omega_j$ to be $\displaystyle k_B T \ln\sinh\left( \frac{\hbar\omega_j}{2k_B T}\right)$, even though it is a complex frequency in general, in calculating the van der Waals force between macroscopic bodies with inhomogeneous dielectric media \cite{parsegian1972van,ninham2003van}, and Barash and Ginzberg suggested using a thermodynamic relations to analyze the electromagnetic fields with matter \cite{barash1975electromagnetic,barash1976expressions,barash1984some}, Schwinger et al. assumed an expression for stress tensor between dielectrics with parallel surfaces for arbitrary temperature to study the Casimir effect in dielectrics, using the methods of source theory \cite{schwinger1978casimir}. My contribution, which is discussed in Ref. \cite{zheng2011lifshitz,zheng2012first}, was to circumvent the difficulties/assumptions inherent in any of the previous generalizations of Lifshitz' theory of van der Waals force and derive a method that relies only on the Maxwell stress tensor for arbitrary electromagnetic fields in vacuum. All calculations of Maxwell stress tensor in my method are restricted to vacuum alone, because of which we don't have to rely on an expression for energy density or stress tensor in dissipative media.

\section{Energy transfer: near-field thermal radiation}

%\subsection{Theory}

Almost all the problems I describe and solve in this manuscript are restricted to planar multilayered media. However, it is rare that we can perform experiments with such objects, or even put to practical use. For instance, in measurements of van der Waals forces, it is common to use spherical or covered objects \cite{tabor1969direct,israelachvili1972measurement,binnig1986atomic,butt1991measuring}. What about the extension of our work to objects of arbitrary shapes? Can that be done using the same theory we used for planar regions - namely to calculate the stress tensor only in vacuum? We do not know the answer to this question though we did embark on the mathematical modeling of electromagnetic fluctuation of arbitrarily shaped objects. Though our work was to get a formalism for momentum transfer between two objects, we succeeded in obtaining a better theoretical result for energy transfer due to the electromagnetic fluctuations.

In the preceding section, we have realized that the fluctuation of electromagnetic waves and fields leads to van der Waals/Casimir force. The electromagnetic waves transport not only momentum, but also energy. Fluctuational electrodynamics can be applied to diverse problems in fluctuation-induced forces as well as radiative energy transfer. The fluctuations of electromagnetic fields lie in the nature of thermal radiation. 

At the beginning of last century, investigations into the fundamental theory and applications of thermal science were very active, however, most were limited to a macroscopic level (large length scales compared to thermal wavelength). Along with the development of micro and nano technology, the small-scale thermal transport became more and more intriguing. During the last few decades, a large number of works were focused on understanding the basic theory of microscopic heat transfer, which has shown a great applications in engineering and military, such as thermophotovoltaics and energy harvesting \cite{qi2010nanotechnology,basu2007microscale,laroche2006near,narayanaswamy2003surface,bernardi2013surface}, manipulation and control of thermal emission \cite{greffet2002coherent,hackermuller2004decoherence,sai2001spectral}, thermal rectification \cite{iizuka2012rectification,zhu2013ultrahigh}, micro/nano device fabrication \cite{basu2007microscale,park2008performance}, bioengineering \cite{rousseau2009radiative,parsegian2006van,yachmenev2004use}, and thermal camouflage and imaging \cite{pantano1997superresolution,novotny1998near,durig1986near,rubevziene2008evaluation}. 

\textit{Why does thermal radiative transfer at microscopic length scale become important?} When undergraduates are taught classical heat transfer, most time devoted to two categories out of three - conductive heat transfer and convective heat transfer. Radiation appears not as important as conduction and convection. It seems you would know radiation quite well if you know Planck's law of blackbody radiation and Stefan-Boltzmann law. That might be true, but only for the far-field region.

As separation between objects gets smaller and smaller, one of the small-size effects begins to dominate - \textit{near-field effect}, which is due to interference, diffraction and tunneling of evanescent waves. If objects are far away, the propagating waves, that are perpendicular to the interface of two bodies, dominate the electromagnetic interaction. At an infinitely large distance in free space, there exists a constant and distance-independent contribution to radiation - blackbody radiation, also known as ``far-field" radiation. There is another type of wave - surface wave (also known as surface phonon or plasmon polariton) - traveling on the surface of material, decaying exponentially along the interface between two media. When the characteristic length of objects $l$ or  the separation $d$ is reduced to be comparable to the thermal wavelength $\lambda$, the contribution due to surface waves dominates the thermal radiation. As compared to blackbody (far-field) radiation, it is called as \textit{near-field} thermal radiation. It has been shown that the near-field thermal radiative heat transfer can exceed Planck's blackbody radiation by several orders of magnitude \cite{hu2008near,narayanaswamy2008thermal,basu2009maximum,zhang2007nano,basu2009review}. %How ``near" is it when the near-field effects are present? It is hard to say because it depends on the absorption spectra and temperature of objects, but I will give a rough approximation of separation of 1 micron or less.

The theory of electric and magnetic fluctuations and thermal radiation proposed by Rytov \cite{rytov1967theory} in 1950s, enabled research on the micro/nano-scale radiative heat transfer. Like Lifshitz' approach of evaluating van der Waals force, the study of radiative heat transfer also requires the spectral characteristics of the optical and emissive properties of materials \cite{zhang2009theory,siegel2002thermal}, which can be manipulated by modifying the properties of materials \cite{lee2006design,maier2005plasmonics,avitzour2009wide} by fabricating micro/nano and/or periodic structures \cite{sai2001spectral,sergeant2010high,lussange2012radiative,sharon1997resonant}. Other efforts of investigating small-scale heat transfer are focused on surface phonon polaritons \cite{shen2009surface,li2013surface,chen2005surface,huber2005near}, surface plasmon polaritons \cite{khurgin2007surface,maier2005plasmonics,kawata2001near}, meta-materials \cite{avitzour2009wide,wu2012metamaterial,salandrino2006far}, photonic bandgaps \cite{knight1998photonic,krauss1996two}, rough surfaces \cite{biehs2010near,fu2009near,smolyaninov1997near}, and nanowires and particles \cite{canetta2013sub,laroche2006near,perez2008heat,maier2002observation}.

Since the study of thermal radiation is tied with the electromagnetic waves and fields that require solving the partial differential equations like Maxwell equation and Helmholtz equation \cite{biehs2010mesoscopic}, a series of computational methods have been performed. These include the finite-differernce-time-domain method \cite{rodriguez2011frequency,wen2010direct}, the numerical scattering method \cite{otey2014fluctuational,bimonte2009scattering,sasihithlu2011proximity}, the molecular dynamics method \cite{domingues2005heat}, the dyadic Green's function approach \cite{narayanaswamy2013green,tai1994dyadic}, and the rigorous coupled-wave analysis \cite{busch2007periodic,grann1996hybrid}, to study the thermal radiative properties of the micro and nano-scaled structures and materials.

Using the dyadic Green's function technique and Rytov's fluctuational electrodynamics, I developed a general formalism for near-field radiative energy and momentum transfer between arbitrarily shaped objects with frequency dependent dielectric permittivity and magnetic permeability in Ref. \cite{narayanaswamy2013green}. It focused on the relation between cross-spectral densities of electromagnetic fields in thermal non-equilibrium which required the evaluation of Poynting vector and electromagnetic stress tensor. The volume integral expressions for cross-spectral densities components of the electric and magnetic fields were obtained, and they can be converted into a form in terms of surface integrals of products of tangential components of the dyadic Green's functions on the surfaces of scatters \cite{narayanaswamy2013green}. The use of a surface integral formalism, replacement of a volume one, can reduce the computational cost dramatically. 

Radiative energy transfer and momentum transfer have the same origins in fluctuational electrodynamics, but they are different in many aspects. We showed energy transfer and fluctuation-induced van der Waals force are qualitatively and quantitatively different due to the disimilar zones of influence of interactions \cite{zheng2014patch}. From the spectral contributions of near-field radiative transfer and van der Waals force, it has been identified the different frequency intervals of interest in the evaluation of the Poynting vector and stress tensor. While much has been learned from the analysis of near-field interactions between two half-spaces separated by a vacuum gap \cite{antezza2008casimir,pitaevskii2006comment,zheng2011lifshitz,kruger2011nonequilibrium,basu2011maximum}, we investigated the surface patch contribution on one of the half-spaces to energy and momentum transfer at any location within the vacuum gap , and showed that contributions from different surface patches are similar for half-spaces with dielectric materials or metals, though their optical properties can be significantly different \cite{palik1998handbook}. The difference is that for energy transfer, all portions of surface contribute positively since energy transfer always takes place from higher to lower temperatures; however for momentum transfer, certian portions of surface contribute to a repulsive force while the rest contributes an attractive force. It may be possible to create objects with net repulsive van der Waals force by truncating or texturing the surfaces appropriately \cite{levin2010casimir,rodriguez2011casimir,maboudian1997critical,israelachvili2011intermolecular}.

\section{Entropy transfer: entropy due to near-field radiative energy transfer}

%\subsection{Theory}

\begin{figure}[h]
\centering
\includegraphics[width =7.0cm]{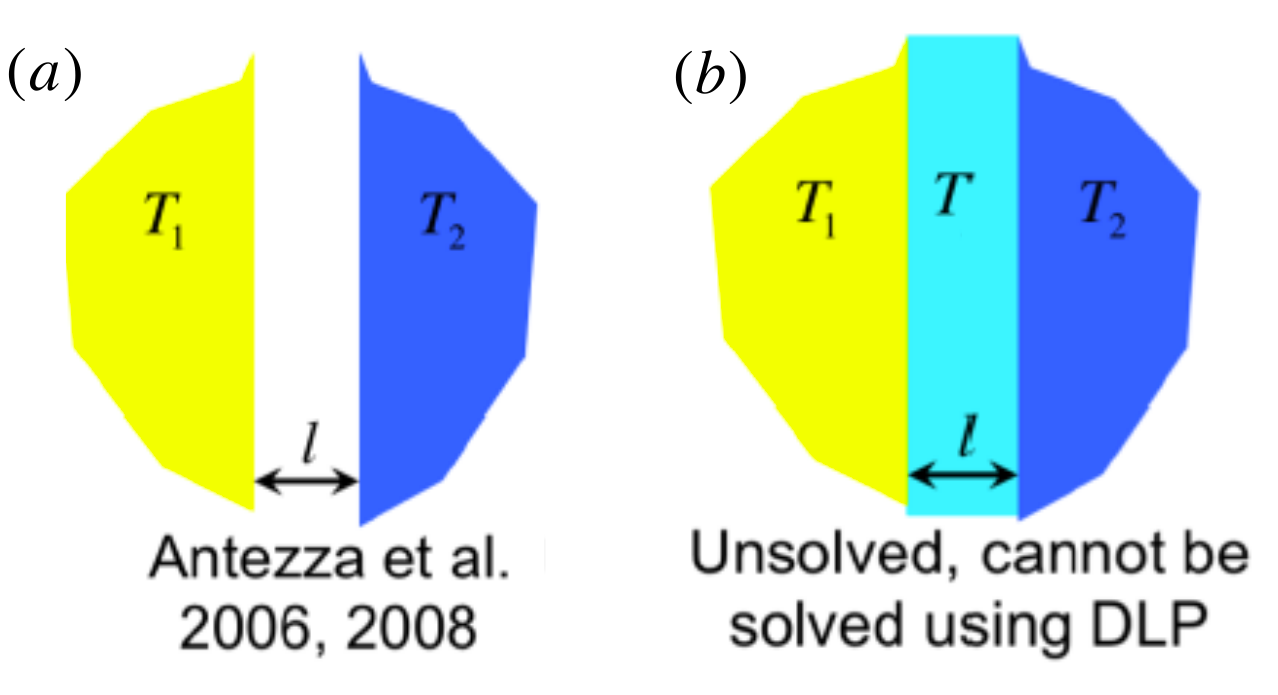}
\caption{\label{fig:neqforce} van der Waals force at thermal non-equilibrium between two half-spaces at temperature $T_1$ and $T_2$ separated by (a) vacuum gap of $l$, that was solved by Antezza et al. in 2006, 2008, and (b) a dissipative medium at temperature $T$, that is an unsolved problem so far.
}
\end{figure}

My main motivation for investigating entropy transfer due to near-field thermal radiation is that I wanted to solve a more complicated and general problem of van der Waals force in any media with dissipative properties. So far, we have a theory of van der Waals force in vacuum at uniform temperature by Lifshitz \cite{lifshitz1956theory} (Fig. \ref{fig:eqforce}(a)), a general theory of van der Waals force in a dissipative medium at thermal equilibrium by Dzyaloshinskii et al. \cite{dzyaloshinskii1961general} and by Zheng and Narayanaswamy \cite{zheng2011lifshitz} (Fig. \ref{fig:eqforce}(b)), and a theory of van der Waals force out of thermal equilibrium valid in vacuum \cite{antezza2005new,antezza2008casimir} (Fig. \ref{fig:neqforce}(a)). However, the van der Waals and/or Casimir force in any dissipative media that is valid at thermal non-equilibrium has not been studied well (Fig. \ref{fig:neqforce}(b)). My work focuses mainly on the momentum transfer. Since we have resolved the problem for van der Waals force in a thin film with dissipative properties, without using any quantum field theory employed by Dzyaloshinskii et al. \cite{dzyaloshinskii1961general}, I devoted myself to a more general problem of van der Waals force out of thermal equilibrium between two half-spaces at temperature $T_1$ and $T_2$ separated by a dissipative medium at temperature $T$, as shown in Fig. \ref{fig:neqforce}(b). Applying the same logic for thermal equilibrium van der Waals force in a dissipative medium to a non-equilibrium case, we came with the existences of singularities and infinite large terms. The singularities have contributions mostly from the properties of dissipative materials, while the infinitely large terms between bodies at different temperatures cannot be simply eliminated by the blackbody radiation or the contribution at an infinite spacing, that led us to perform a thermodynamic analysis for near-field thermal radiation and to study the thermal non-equilibrium entropy transfer at micro/nano length scale.

In the preceding section, radiative heat transfer is the key to analyzing the solar energy conversion and thermophotovolatics \cite{laroche2006near,zhang2008radiation}. It has been shown that it can be enhanced by several orders of magnitude at small separations, as compared to the blackbody limit, and it can be used to improve the energy conversion efficiency and performance of the macroscopic and microscopic devices such as solar cells, polariton assisted nano-lithography masks, scanning tunneling microscopes. 

The current study of thermal radiation can be traced back to Planck's pioneering work on blackbody radiation more than a century ago \cite{planck2011theory,siegel2002thermal}. He applied thermodynamic analysis of radiation in a vacuum cavity, which required the knowledge of energy, momentum and entropy of photons. Planck was not first to investigate the thermodynamics of radiation. In 1884, Boltzmann focused on an isothermal enclosure at temperature $T$ and derived a well-known formula for blackbody emissive power $e_b(T)=\sigma T^4$, where $\sigma$ is the Stefan-Boltzmann constant \cite{blevin1971precise,campisi2005mechanical}. Then the associated entropy power was determined to be $s_b(T)=\displaystyle \frac{4}{3} \sigma T^3$. In the 1900s, Planck introduced quanta to radiation, and established Planck's law of blackbody radiation while expressing the spectral radiative energy as well as entropy intensity of a monochromatic plane polarized ray of frequency $\nu$ \cite{planck2011theory}.

Planck's contribution is pioneering and impressive, but his work is restricted to the case when near-field effects are absent. Within the past century, some efforts were made for entropy or availability or usable work. In 1910s, von Laue studied the entropy of the interfering eletromagnetic beams. In 1964, Petela introduced exergy of radiation to energy conversion. Landsberg and Tonge \cite{landsberg1980thermodynamic}, Jeter et al. \cite{jeter1981maximum}, and Gribik et al. \cite{gribik1984second} introduced the concepts of dilute blackbody radiation and effective temperature to evaluate the maximum work that can be extracted from solar radiation in the 1980s. In 1983, Brakat and Brosseau investigated the entropy of N partially coherent pencils of radiation. The entropy of far-field thermal radiation has been studied extensively in the second half of last century and in the first decade of this century \cite{rueda1973entropy,ladd1982energy,kabelac2008thermodynamic,perez2011mesoscopic,kruger2011nonequilibrium,landsberg1980thermodynamic,zhang2007entropy,park2008performance,kabelac2012entropy,latella2014near}.
For example, the entropy of graybody radiation was calculated \cite{wright2004exergetic}; the near-field radiation and far-field radiation were compared by extending the far-field theory to near-field concept without taking near-field effects into consideration actually \cite{whale2001effective}, the radiative entropy generation was studied for the participating media between blackbody walls while neglecting the entropy generation at the walls, which could be important \cite{caldas2005entropy,liu2006entropy}. 

In 2007, Zhang \cite{zhang2007entropy,zhang2008radiation} studied the entropy associated with reflection, emission and transmission of thermal radiation by a surface, and derived for entropy generation during radiative transfer between isothermal diffuse-gray surfaces. Zhang's work was still in the far-field limit, for which the near-field effects were not mentioned or touched yet. In 2011, Dorofeyev evaluated the position dependent energy and entropy density while considering the position dependent (local) density of states at thermal equilibrium \cite{dorofeyev2011thermodynamic}. The work in Ref. \cite{narayanaswamy2013theory} investigated the near-field radiative entropy density and entropy flux that is valid for both thermal non-equilibrium as well thermal equilibrium cases, and it shows an agreement with the theory of blackbody radiation in the far-field limit, due to multiple reflections, interference and diffraction of light \cite{narayanaswamy2013theory,latella2014near,sokolsky2014scaling,butt2014entropy}.

%\subsection{Measurements and Applications}

Radiative entropy is a measure of unavailable work that cannot be extracted from thermal radiation. The goal is to improve the performance of energy conversion systems, namely, increasing the amount of useful work and reducing the unavailable work. For example, several measurements of radiative entropy have been performed for the thermal and quantum noise \cite{oliver1965thermal}, for quantum optical correlation \cite{barnett1989entropy}, for partially polarized radiation, and its application to estimating radio sky polarization distributions \cite{ponsonby1973entropy}. Indirect measurement for near-field radiative entropy is to measure thermal quantities such as radiative energy transfer, available work, and energy conversion efficiency, i.e., thermal effiiciency and quantum efficiency, which are relatively easier to measure. For example, the measurement for radiative heat transfer can be found in Refs. \cite{shen2009surface,narayanaswamy2011heat,gu2013experimental,fu2009thermal,hu2008near,joulain2003definition,everson1990measurement}  . It is not accurate to determine entropy while using the measured conversion efficiency of energy \cite{park2008performance,ruan2007entropy,latella2014near,parrott1978theoretical}, which is in general much lower than the theoretical thermodynamic conversion efficiency \cite{park2008performance,narayanaswamy2013theory}.
%or local density of states %\cite{meyer2003direct,joulain2003definition,everson1990measurement,krachmalnicoff2010fluctuations}.
The knowledge of radiative entropy transfer between macroscopic objects enables us to determine the maximum work through thermal radiation due to near-field effects \cite{basu2009maximum,narayanaswamy2013theory}, to design high efficienty energy conversion systems for energy harvesting \cite{park2008performance,basu2007microscale,latella2014near}, to understand the thermodynamics of surface wave-based laser cooling \cite{khurgin2007surface}, to study the thermal and entropic contributions to non-equilibrium van der Waals/Casimir interactions (not fully developed yet) \cite{narayanaswamy2013van}, and to utilize solar power \cite{parrott1978theoretical,zhang2008radiation,gribik1984second,boehm1986maximum,sergeant2010high}.

\section{Summary}

This manuscript focuses on the transport of energy, momentum and entropy due to electromagnetic fluctuations with near-field effects taken into consideration. For momentum transfer, I give a new perspective to the theory of van der Waals pressure by obtaining the results of Dzyaloshinskii, Lifshitz, and Pistaevskii without having to use any quantum field theory. I show that the computation of van der Waals pressure between objects on the imaginary frequency axis is only a numerical/mathematical convenience, not a physical necessity. For energy transfer, I identify some of the similarities and differences between energy and momentum transfer. I solve a problem in near-field radiative transfer between two half-spaces to identify the differences, mainly with an aim of identifying features that make it likely that the proximity approximation for computing near-field radiative transfer between two curved objects is as valid as the proximity approximation for van der Waals forces between curved surfaces. The analysis shows qualitative differences between energy and momentum transfer. Finally, I solve for the first time the entropy transfer between half-spaces at different temperatures taking near-field effects into account. I wanted to calculate the momentum and entropy transfer between two half-spaces in order to solve the more complicated problem of van der Waals pressure in a layer of dissipative material between two half-spaces at different temperatures, namely the problem of Dzyaloshinskii, Lifshitz, and Pitaevskii but under conditions of thermal non-equilibrium. My hypothesis was that the knowledge of non-equilibrium entropy transfer in a vacuum gap would furnish us the solution. I have not been successful in that endeavor, though. This work is devoted to establishing a general theory of momentum, energy and entropy transport between arbitrarily shaped objects at thermal non-equilibrium and at a microscopic length scale, which urges a more careful, deeper, and complete thermodynamic study of near-field radiative heat transfer.

\newpage
\section*{References}
\bibliography{YiZhengBib}
%
%%\begin{thebibliography}{10}
%%\bibitem{ref1} J.~Doe, Article name, \textit{Phys. Rev. Lett.}
%%
%%\bibitem{ref2} J.~Doe, J. Smith, Other article name, \textit{Phys. Rev. Lett.}
%%
%%\bibitem{web} \href{http://www.google.pl}{www.google.pl}
%%\end{thebibliography}

\end{document}